      [1999/12/01 v1.4c Il Nuovo Cimento]
\documentclass{cimento}

\newcommand{\sqrts}{\sqrt{s}}
\newcommand{\sqrtsNN}{\sqrt{s_{\scriptscriptstyle \rm NN}}}

\newcommand{\mev}{\mathrm{MeV}}
\newcommand{\gev}{\mathrm{GeV}}
\newcommand{\tev}{\mathrm{TeV}}
\newcommand{\fm}{\mathrm{fm}}

\newcommand{\cm}{\mathrm{cm}}

\newcommand{\pt}{p_{\rm t}}

\usepackage{graphicx}  % got figures? uncomment this
\title{First ALICE results from heavy-ion collisions at the LHC}
\author{A.~Dainese~\from{ins:x}\thanks{e-mail: andrea.dainese@pd.infn.it}, for the ALICE Collaboration}
\instlist{\inst{ins:x} INFN -- Sezione di Padova, Padova, Italy}
\PACSes{\PACSit{24.85.+p, 25.75.-q, 25.75.Ag}{}
}
\begin{document}

\maketitle

\begin{abstract}
The ALICE detector recorded Pb--Pb collisions at $\sqrtsNN = 2.76$~TeV at the LHC
in November--December 2010. We present the results of the measurements that provide a
first characterization of the hot and dense state of strongly-interacting matter
produced in heavy-ion collisions at these energies. 
In particular, we describe the measurements of the
particle multiplicity, collective flow, Bose-Einstein correlations, high-momentum 
suppression, and their dependence on the collision centrality. These observables
are related to the energy density, the size, the viscosity, and the opacity of the
system. Finally, we give an outlook on the upcoming results, with emphasis on
heavy flavour production.
\end{abstract}

\section{Introduction}
\label{sec:intro}

The ALICE experiment~\cite{aliceJINST} studies nucleus--nucleus and 
proton--proton collisions at the Large Hadron Collider, with the
main goal of investigating the properties of the high-density 
state of QCD matter that is expected to be formed
in Pb--Pb collisions~\cite{PPRv1,PPRv2}. According to lattice QCD 
calculations, under the conditions of high energy density and 
temperature reached in these collisions, the phase transition to a Quark-Gluon Plasma (QGP)
would occur, colour confinement of quarks and gluons 
into hadrons would be removed and chiral symmetry would be restored (see e.g.~\cite{karsch}). 

The ALICE detector was designed in order to provide tracking and particle identification
over a large range of momenta
(from tens of MeV/$c$ to over 100~GeV/$c$), low material budget and excellent vertexing capabilities. These features have been tailored to reach a
detailed characterization of the state of matter produced in Pb--Pb collisions, with particular attention to global event
properties and hard probes. 

\begin{figure}[!t]
\begin{center}
\includegraphics[width=0.49\textwidth]{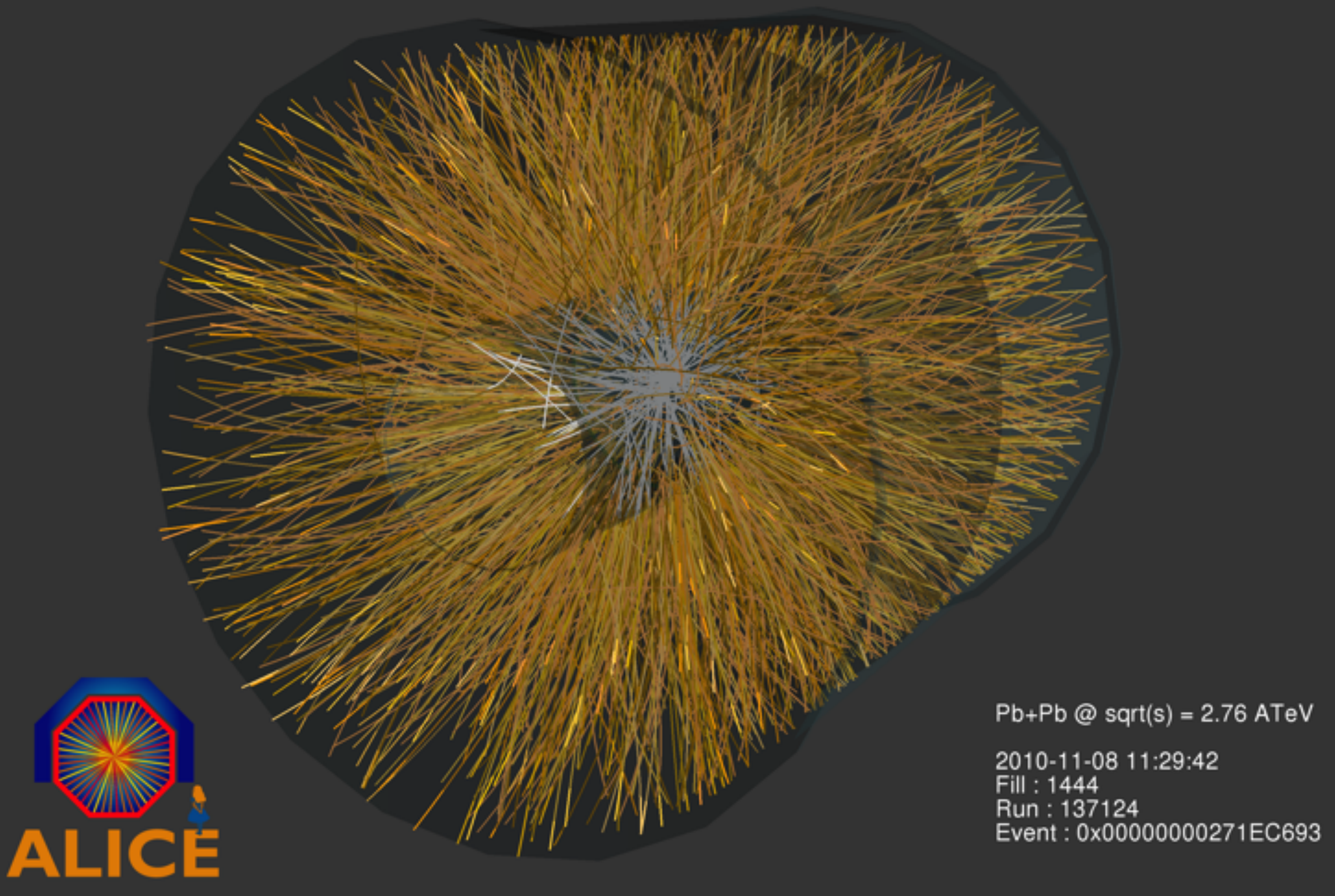}
\includegraphics[width=0.49\textwidth]{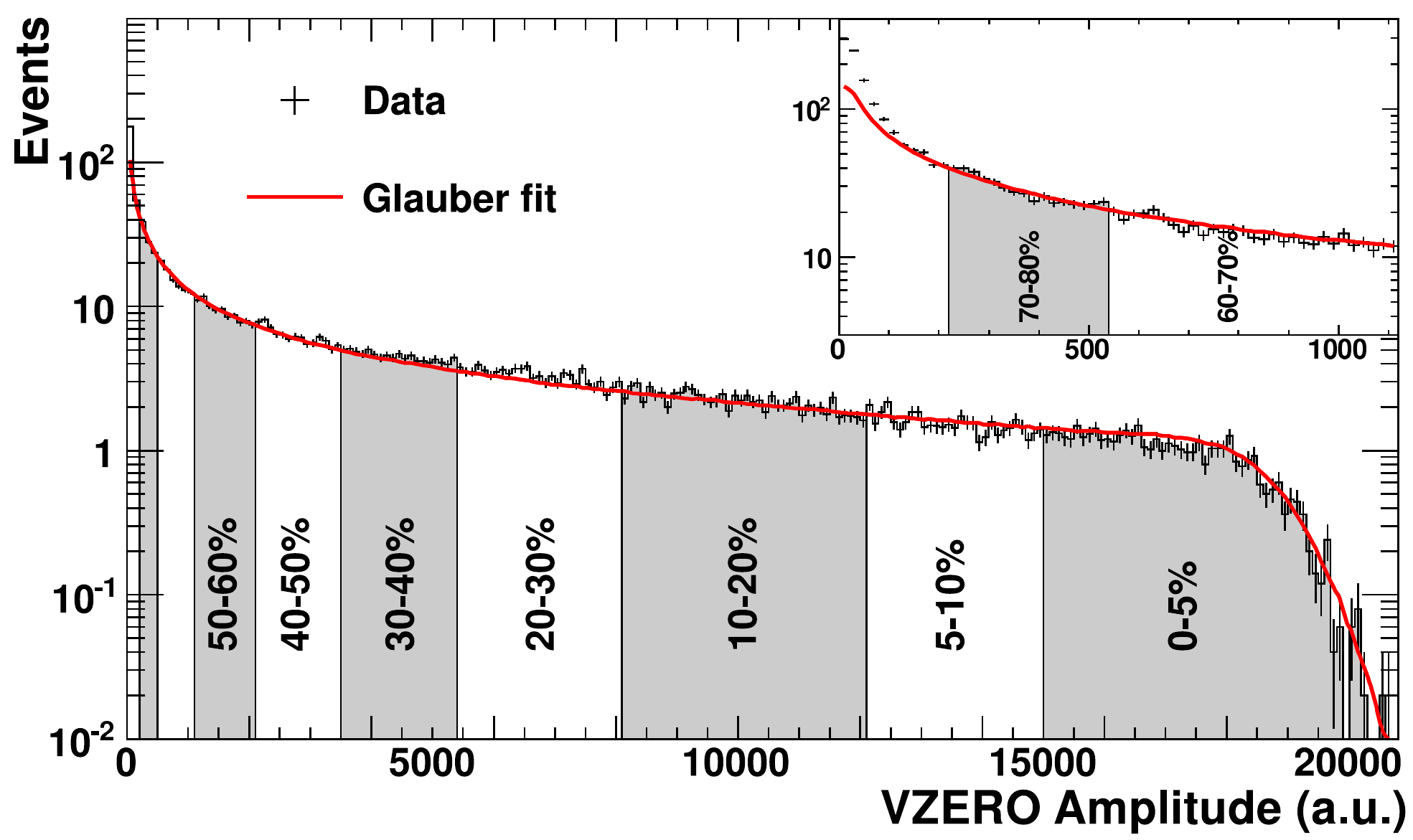}
\caption{Left: tracks reconstructed in the ALICE Time Projection Chamber and Inner Tracking System in one of the first Pb--Pb collisions recorded by the detector.
Right: distribution of the summed amplitudes in the VZERO scintillator tiles (histogram); 
inset shows the low amplitude part of the distribution; 
the curve shows the result of the Glauber model fit to the measurement. 
the vertical lines separate the centrality classes used in the analysis~\cite{cent}.}
\label{fig:display}
\end{center}
\end{figure}

The experiment has collected the first Pb--Pb data in November--December 2010 at 
a centre-of-mass energy $\sqrtsNN=2.76~\tev$ per nucleon--nucleon collision (see 
Fig.~\ref{fig:display}, left). With a 
fourteen-fold increase with respect to nucleus--nucleus collisions at the RHIC collider
(Au--Au at $\sqrtsNN=200~\gev$), this constitutes the largest energy increase in the history of 
heavy-ion Physics and, as such, it opens new exciting scenarios for the study of high-density 
QCD matter. During the Pb--Pb run and shortly after it, the first results on the characterization 
of this state of matter were obtained~\cite{mult,flow,cent,hbt,raa}. These results are summarized
in the present report. 

In section~\ref{sec:detector}, the ALICE experimental setup is briefly
described, with emphasis on the detectors that were used for the results presented here,
along with the  data collection and collision centrality determination. 
The most fundamental measurement
that characterizes the inclusive particle production is reported in section~\ref{sec:mult}: 
the charged particle multiplicity density~\cite{mult} and its dependence on the 
collision centrality~\cite{cent}. This measurements provides information on the 
energy density of the system and, via comparison with models, on the gluon dynamics in 
the high-energy colliding nuclei.
In section~\ref{sec:flow} the elliptic flow measurement is described, compared to 
lower-energy data, and related to the hydro-dynamical properties of the 
produced system~\cite{flow}.
In section~\ref{sec:hbt} the measurement of the 
Bose-Einstein two-pion correlation, that allows to characterize the spatial extension 
of the particle emitting source, is described~\cite{hbt}. 
The study of the suppression of the charged particle production at large momentum,
via the so-called nuclear modification factor, is presented in section~\ref{sec:raa}.
Finally, in section~\ref{sec:outlook},
an outlook is given on the ongoing analyses, 
which will provide further insight on the 
QCD medium properties.

\section{ALICE detector, Pb--Pb data sample, and collision-centrality determination}
\label{sec:detector}

The ALICE apparatus is described in~\cite{aliceJINST}. It consists of two main parts: 
a central detector,
placed inside a solenoidal magnet providing a field of up to 0.5~T, where charged and 
neutral particles
are reconstructed and identified in the pseudorapidity range $|\eta|<0.9$,
and a forward muon spectrometer covering the range $-4<\eta<-2.5$. The apparatus is completed
by a set of smaller detectors in the forward areas, for triggering, charged particle and photon 
counting, and event classification.

The main results presented in this report were obtained using the following ALICE detectors:
the VZERO scintillators, the Inner Tracking System (ITS), the Time Projection Chamber (TPC).

The two forward scintillator hodoscopes (VZERO) are segmented into 32 scintillator counters each, arranged in four rings around the beam pipe. They  cover the pseudorapidity ranges $2.8 < \eta < 5.1$ and $-3.7 < \eta < -1.7$, respectively. 
The ITS is composed of high resolution silicon tracking detectors, arranged in six cylindrical layers at radial distances to the beam line from 3.9 to 43 cm. Three different technologies are employed:
Silicon Pixel Detectors (SPD) for the two innermost layers, Silicon Drift Detector (SDD) 
for the two intermediate layers, and Silicon Strip Detector (SSD) for the two outermost layers. 
The TPC is a large cylindrical drift detector
with cathode pad readout multi-wire proportional chambers at the two edges.
 The active volume is $85 < r < 247$~cm and $-250 < z < 250$~cm in the radial and longitudinal directions, respectively. 

All data presented in this report were collected with a magnetic field of 0.5~T and  
a minimum-bias trigger requiring at least two out these three conditions:
a hit in the SPD, a hit in the forward rapidity VZERO counters, or a hit in the backward rapidity 
VZERO counters. This request selects about 98\% of the Pb--Pb inelastic cross section.
The instantaneous luminosity was typically of the order of $10^{25}~\rm cm^{-2}s^{-1}$ during
the Pb--Pb run and a total statistics of about 30 million minimum-bias triggers was recorded,
in addition to high-multiplicity and ultra-peripheral collision triggers.

Nucleus--nucleus collisions are classified according to their centrality, which measures the 
number of nucleons that undergo inelastic scattering (number of participants, $N_{\rm part}$),
and is related to the initial extension of the system produced in the collision. Several experimental
observables, mainly measures of the number of particles produced in the collisions, 
can be used to categorize the events in centrality classes. Figure~\ref{fig:display} (right) shows the 
distribution of the observable that was used for the first analyses of Pb--Pb data collected 
in 2010~\cite{cent}:
the sum of amplitudes in the VZERO scintillator detector, the response of which 
is proportional to the event multiplicity.
The distribution is fit using the Glauber model~\cite{glauber} to describe the collision geometry and a Negative Binomial Distribution (NBD) to describe particle production~\cite{nbd}. 
In addition to the two parameters of the NBD, there is one free parameter that controls the power-law dependence of particle production on the number of participating nucleons 
($N_{\rm part}$). 
 The fit is restricted to amplitudes above a value corresponding to 88\% of the hadronic cross section. In this region the trigger and event selection are fully efficient, and the contamination by electromagnetic processes is negligible. Centrality classes are determined by integrating the measured distribution above the cut, as shown in Fig.~\ref{fig:display} (right).

\section{Charged particle multiplicity and its collision-centrality dependence: a high-density 
system from gluon-saturated colliding nuclei}
\label{sec:mult}

The multiplicity of charged particles per unit of pseudo-rapidity ($\eta$) at central rapidity 
is measured using the Silicon Pixel Detector, the 
innermost sub-detector of the Inner Tracking System, made of two layers with radii of 
$3.9$ and $7.6~\cm$, and with acceptances $|\eta|<2.0$ and $|\eta|<1.4$, respectively.  
Tracklet candidates are formed using information on the position of the primary vertex,
reconstructed with the same detector, and of hits on the two layers. In particular,
a tracklet is defined by a pair of hits, one on each layer, selected on the basis of their
polar and azimuthal angles, so that the resulting tracklet points to the primary vertex.
The cut imposed on the azimuthal angle efficiently selects charged particles with transverse momentum ($\pt$) above $50~\mev/c$. 
Particles below $50~\mev/c$ are mostly absorbed by material.
The charged-particle pseudo-rapidity density ${\rm d} N_{\rm ch} /{\rm d}\eta$ is obtained from the number of tracklets within $|\eta| < 0.5$, corrected for acceptance, efficiency and background
contamination. The background is estimated from the data and from simulations with three
different methods~\cite{mult}.

\begin{figure}[t!]
\begin{center}
\includegraphics[width=0.49\textwidth]{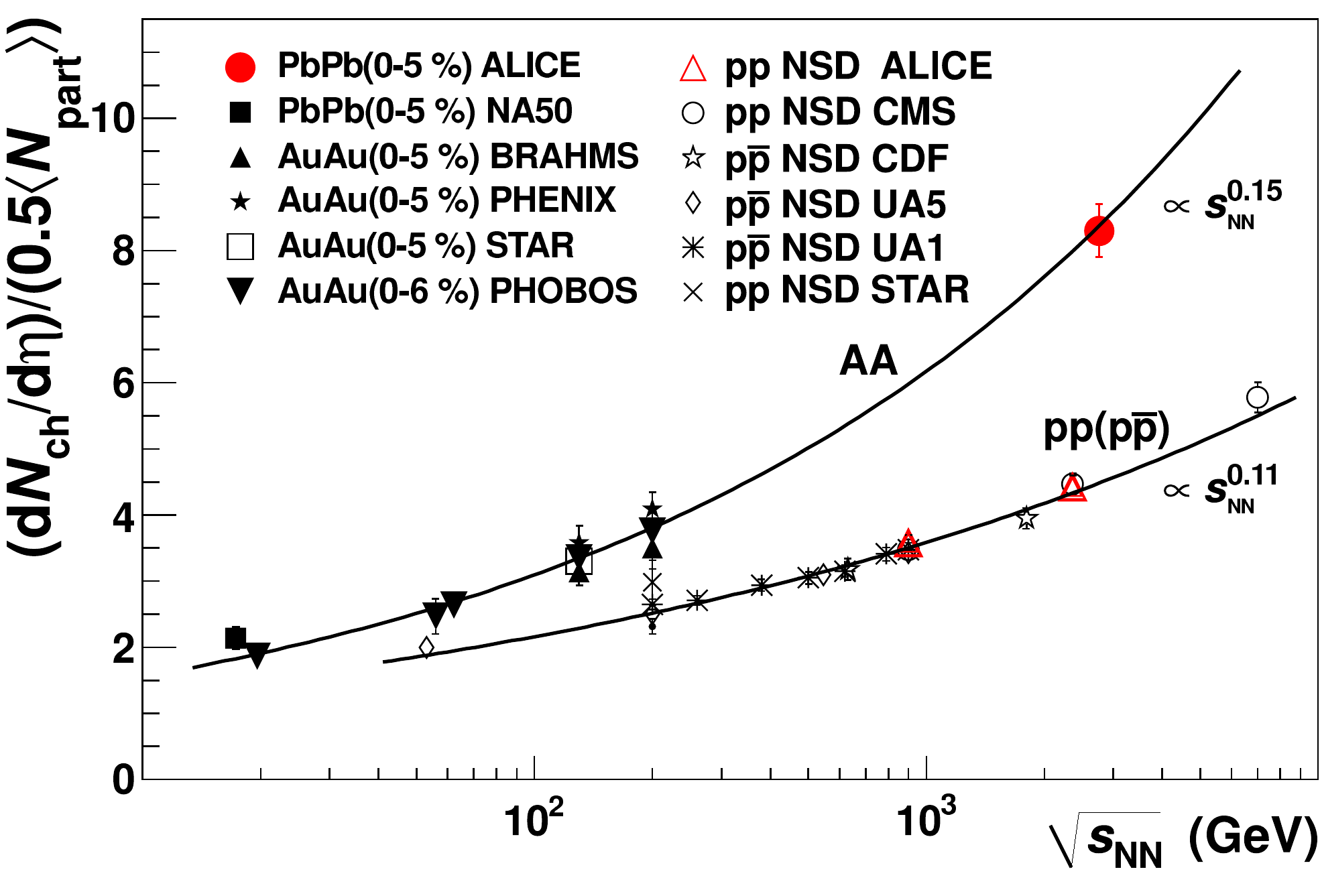}
\includegraphics[width=0.49\textwidth]{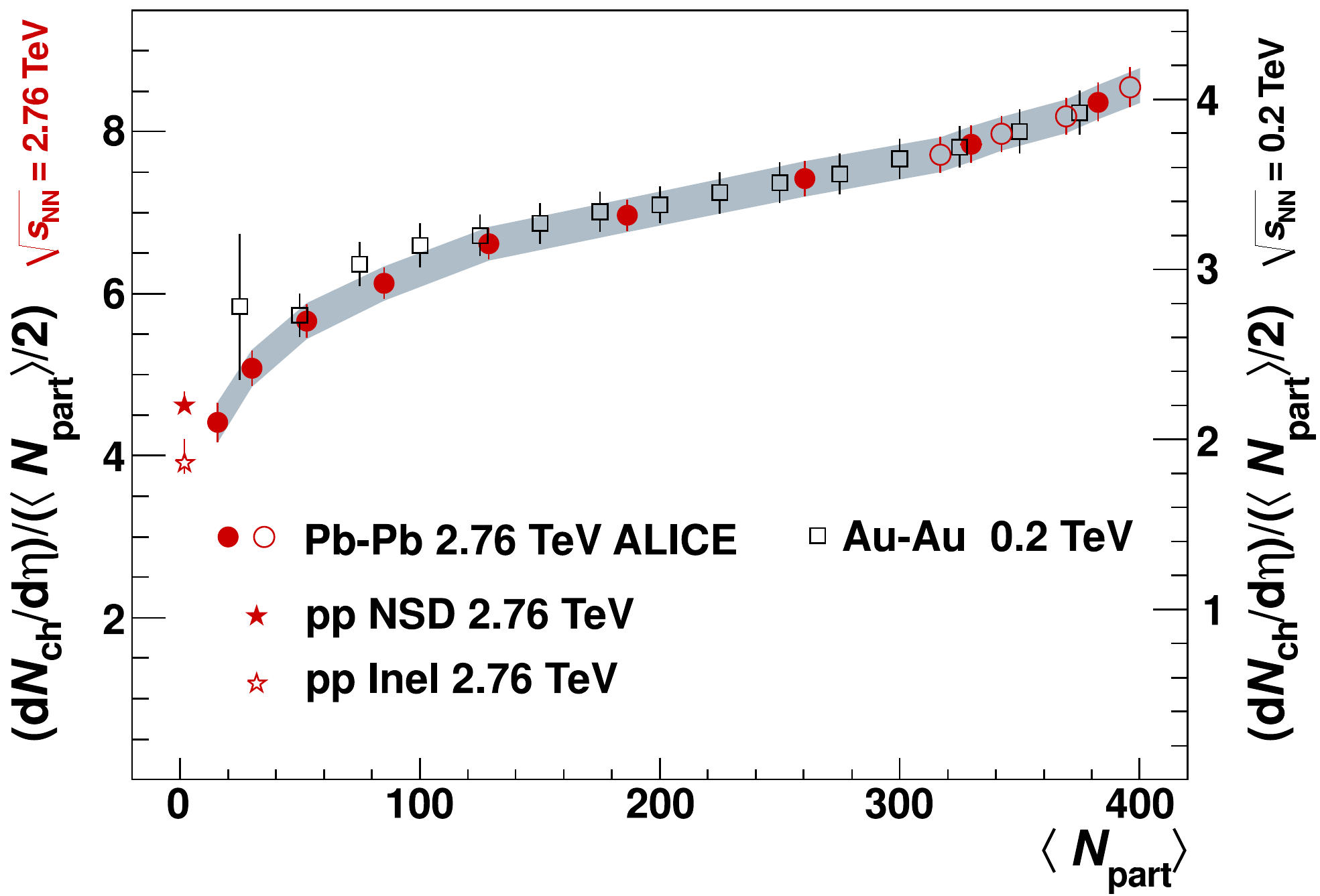}
\caption{\label{fig:mult} Charged particle multiplicity.
Left: ${\rm d} N_{\rm ch} /{\rm d}\eta$
 per participant pair for central nucleus--nucleus 
and non-single diffractive pp ($\rm p\overline{p}$) collisions, as a function of 
$\sqrtsNN$ (\cite{mult} and references therein). 
Right: the same observable as a function of collision centrality for nucleus--nucleus collisions at 
$\sqrtsNN=2.76$~\cite{cent} and $0.2~\tev$~\cite{phobos} (scaled up by a factor 2.1).
}
\end{center}
\end{figure}

In the 5\% most central Pb--Pb collisions, we measured a density of primary charged particles
at mid-rapidity ${\rm d} N_{\rm ch} /{\rm d}\eta = 1584 \pm 4 (stat.) \pm 76
(sys.)$~\cite{mult}. Normalizing per participant pair (using $N_{\rm part}$ from the Glauber model fit), 
we obtained
${\rm d} N_{\rm ch} /{\rm d}\eta /(0.5\langle N_{\rm part}\rangle) = 8.3 \pm 0.4 (sys.)$ 
with negligible statistical error. In Fig.~\ref{fig:mult} (left), this value is compared
to the measurements for Au--Au and Pb--Pb, and non-single diffractive (NSD) 
pp and $\rm p\overline{p}$ collisions over a wide
range of collision energies. It is interesting to
note that the energy dependence is steeper for heavy-ion
collisions than for pp collisions.  A significant increase, by a factor 2.1, in the pseudo-rapidity density is observed at $\sqrtsNN = 2.76$~TeV for Pb--Pb compared to $\sqrtsNN= 200$~GeV 
for Au--Au. Bjorken's estimation of the initial energy density in the system formed in the
collisions reads: $\epsilon = Energy/Volume = {\rm d}N/{\rm d} y \cdot \langle m_{\rm t}\rangle / (\tau_0 \rm A)$, where ${\rm d}N/{\rm d} y$ and $\langle m_{\rm t}\rangle$ are the rapidity density  and the average transverse mass of the produced particles, 
$\tau_0$ is the formation time of the system, and A is the mass number
of the colliding nuclei, which estimates the transverse area of the nuclear overlap for central 
collisions. This relation and our measurement suggest that  the energy density of the 
system produced at LHC energies is at least a factor of 3 larger than at RHIC energies, considering
the 2.1-fold larger multiplicity and the fact that the formation time $\tau_0$ is expected to be
shorter by a factor of about two with respect to RHIC energies.
Figure~\ref{fig:mult} (right) shows the centrality dependence of the charged multiplicity
per participant pair~\cite{cent}, compared to the corresponding RHIC 
measurement~\cite{phobos}, scaled by a factor 2.1. The trend is very similar at the two energies
and the mild increase in semi-central to central collisions is found to be better described by 
models that include a mechanism to tame the increase with centrality in the number of scattering
centres. This suggests a certain degree of saturation in the phase-space of small $x$ 
(fractional momentum) 
gluons in the initial state of the collision.

\section{Elliptic flow: the perfect liquid at the LHC}
\label{sec:flow}

One of the experimental observables that is sensitive to the properties of high-density QCD matter is the azimuthal distribution of particles in the plane perpendicular to the beam direction. In non-central collisions, the geometrical overlap region and therefore the initial matter distribution is anisotropic (almond shaped). If the matter is interacting, this spatial asymmetry is converted via multiple collisions into an anisotropic momentum distribution. This anisotropy is 
quantified via the elliptic flow coefficient, $v_2$, defined as the second moment of the final state hadron azimuthal distribution, ${\rm d}N/{\rm d}\phi$, with respect to the 
reaction plane, which contains the centres of the colliding nuclei and the beam line.
The elliptic flow is a response of the dense system to the initial conditions and therefore 
it is sensitive to the early and hot, strongly interacting phase of the evolution. 
The large elliptic flow measured for Au--Au collisions at RHIC is well-reproduced by models based
on relativistic hydrodynamics with a QGP equation of state and small, but non-zero, viscosity.

\begin{figure}[t!]
\begin{center}
\includegraphics[width=0.51\textwidth]{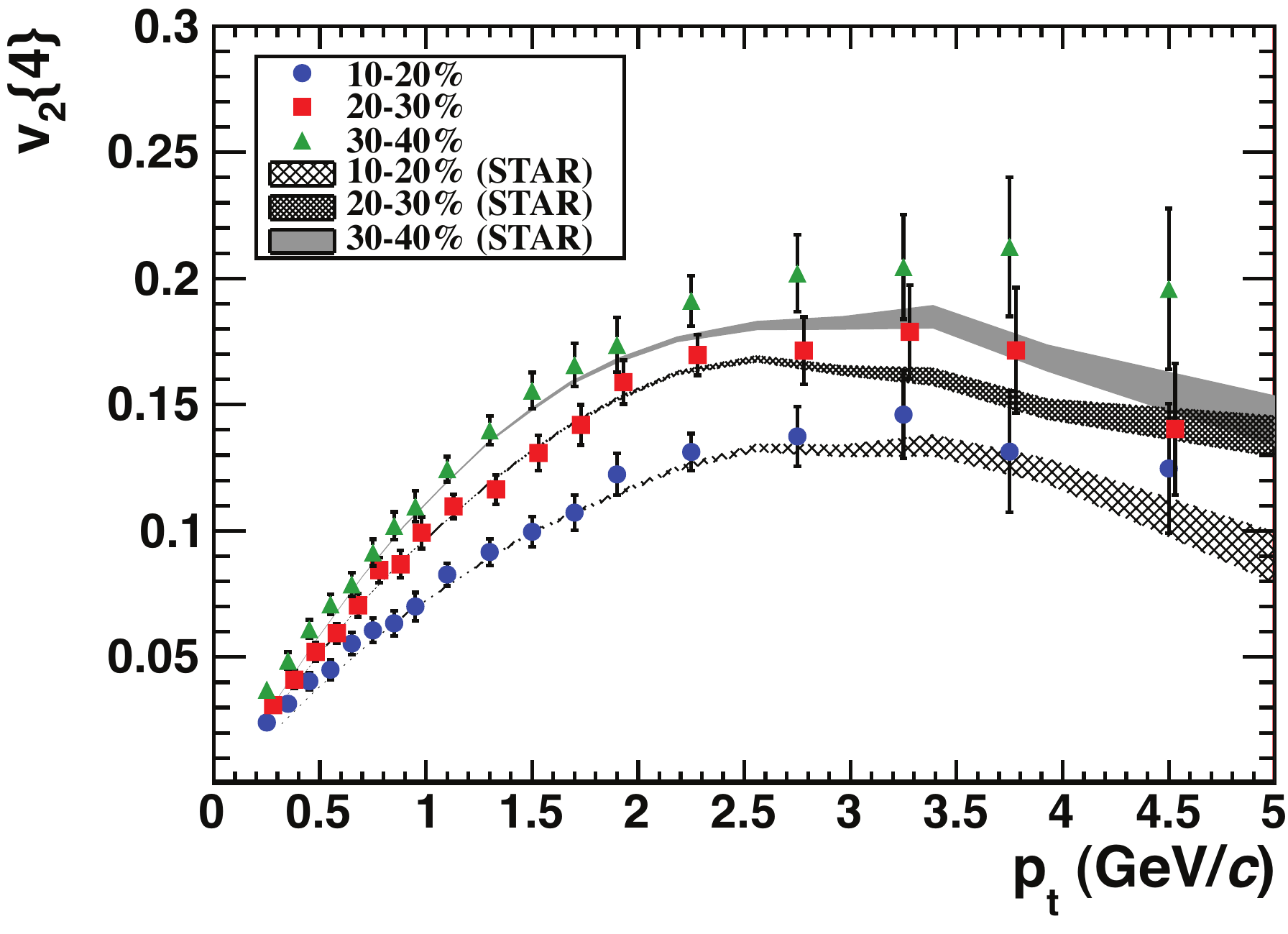}
\includegraphics[width=0.48\textwidth]{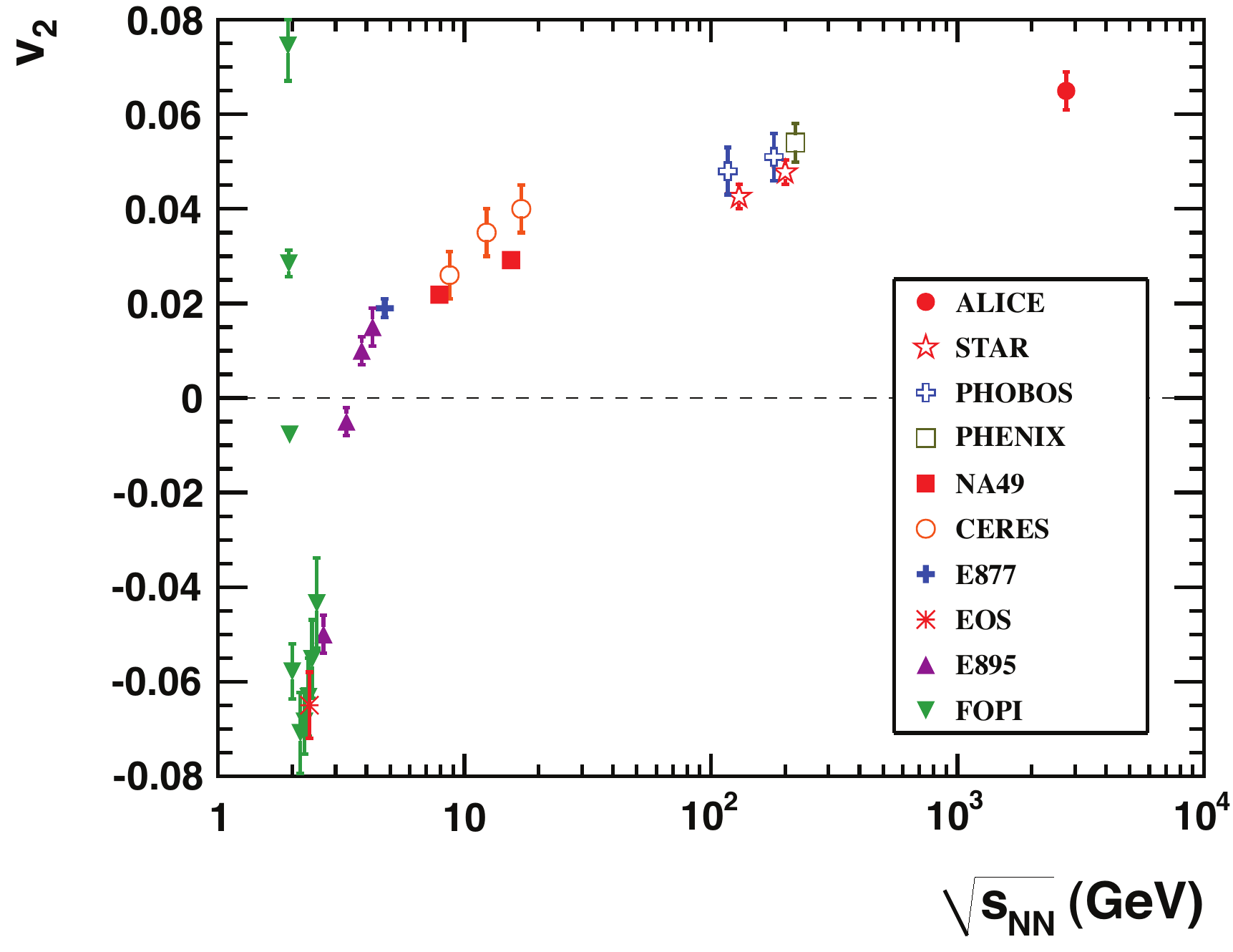}
\caption{\label{fig:flow} Elliptic flow.
Left: $v_2\{4\}(\pt)$ for various centralities compared to STAR measurements. 
Right: Integrated elliptic flow at 2.76 TeV in Pb--Pb 20--30\% centrality class compared with results from lower energies taken at similar centralities~(\cite{flow} and references therein).}
\end{center}
\end{figure}

The first results on the elliptic flow in Pb--Pb collisions at the LHC were obtained using 
charged particle tracks reconstructed in the TPC and in the ITS. The tracks were required to have at least 70 reconstructed space points out of the maximum 159 in the TPC and a $\chi^2$ per TPC cluster $\le 4$ (with two degrees of freedom per cluster). Additionally, at least two of the six ITS layers must have a hit associated with the track, including at least one of the two pixel layers.
A selection based on the distance of closest approach to the primary vertex was used to 
reject a large fraction of the tracks produced by secondary particles, from decays and
interactions in the detector material.
The $\pt$-differential flow was measured for different event centralities using various analysis techniques~\cite{flow}, based on multi-particle cumulants ($v_2\{2\}$ and $v_2\{4\}$).
Figure~\ref{fig:flow} (left) presents $v_2(\pt)$ obtained with the 4-particle cumulant method for three different centralities, compared to STAR measurements at RHIC. 
The transverse momentum dependence is qualitatively similar for all three centrality classes. 
The observed similarity at
RHIC and the LHC of the $\pt$-differential elliptic flow at low $\pt$ is consistent with predictions of hydrodynamic models.
The integrated elliptic flow measured in the 20--30\% centrality class is compared to results from lower energies in Fig.~\ref{fig:flow} (right). The figure shows that there is a continuous increase in the magnitude of the elliptic flow for this centrality region from RHIC to LHC energies.
We find that the integrated elliptic flow increases by about 30\% from $\sqrtsNN= 200$~GeV 
at RHIC to  2.76~TeV. This increase is higher than current predictions from ideal hydrodynamic models. The hydrodynamic models which incorporate viscous corrections and certain hybrid models do allow for such an increase. In these models the increase is due to the reduced 
importance of viscous corrections at LHC energies. This is a first indication that the high-density 
QCD matter produced in Pb--Pb collisions at the LHC resembles closely a perfect liquid,
with close to zero viscosity.

\section{Femptoscopic study: a larger and longer-lived particle emitting source}
\label{sec:hbt}

\begin{figure}[t!]
\begin{center}
\includegraphics[width=0.49\textwidth]{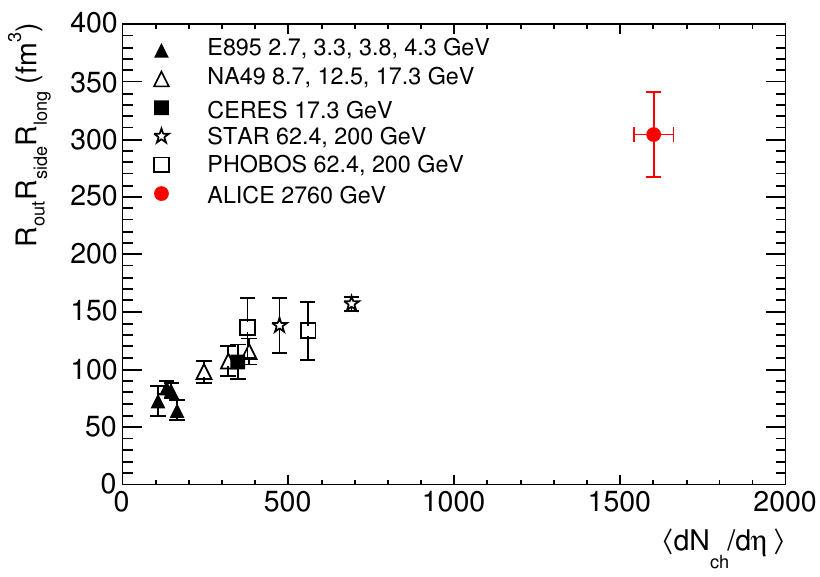}
\includegraphics[width=0.49\textwidth]{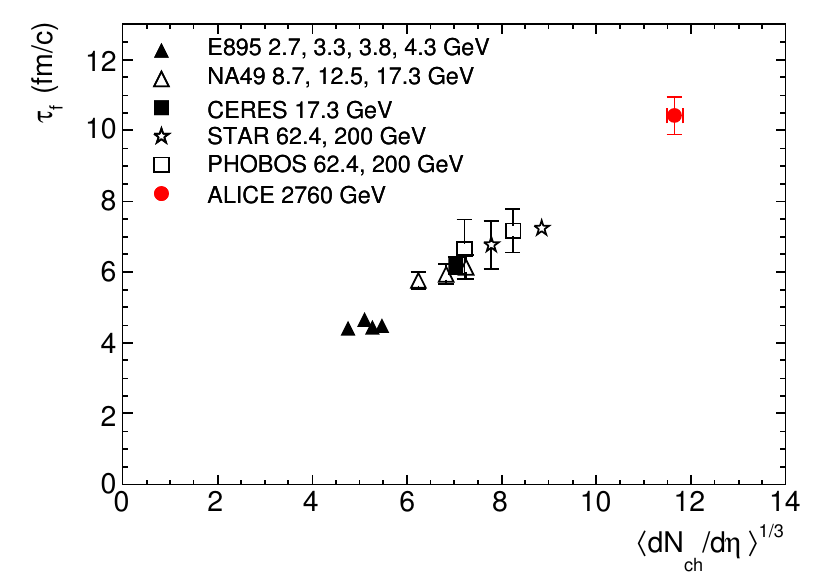}
\caption{Femptoscopic measurements.
Product of the three pion HBT radii (left) and decoupling time extracted from $R_{\rm long}$ (right). The ALICE results (red filled dots) are compared to those obtained for central gold and lead collisions at lower energies at the AGS, SPS, and RHIC. See~\cite{hbt} and references therein.}
\label{fig:hbt} 
\end{center}
\end{figure}

The Bose-Einstein enhancement of identical-pion pairs at low relative momentum allow 
to assess the spatial scale of the emitting source in $\rm e^+e^-$, hadron--hadron, lepton--hadron, 
and heavy-ion collisions. Especially in the latter case, this technique, known as 
Hanbury Brown-Twiss (HBT) interferometry and being a special case of femtoscopy, has been developed into a precision tool to probe the dynamically-generated geometry of the emitting system. See~\cite{hbt} for more details and references. 

The first measurement of the HBT radii for Pb--Pb collisions at the LHC~\cite{hbt} was 
carried out using pion tracks, reconstructed in the TPC and ITS 
(similar selection cuts as for the flow analysis) and identified using the TPC 
specific energy deposit ${\rm d} E/{\rm d}x$. The details on the construction of the
two-pion correlation functions and their analysis are described in~\cite{hbt}.
Figure~\ref{fig:hbt} (left) shows the dependence on charged particle multiplicity of the product of the three HBT radii ($R_{\rm out}\cdot R_{\rm long} \cdot R_{\rm side}$), extracted as the Gaussian widths of the correlation function in three perpendicular directions. 
This product is connected to the volume of the homogeneity region.
In central Pb--Pb collisions 
at $\sqrtsNN=2.76~\tev$ we measured a product of about $300~\fm^3$, about 
two times larger than at RHIC energy.

Within hydrodynamic scenarios, the decoupling time for hadrons at mid-rapidity can be estimated as follows: the size of the homogeneity region is inversely proportional to the velocity gradient of the expanding system; the longitudinal velocity gradient in a high energy nuclear collision decreases with time as $1/\tau$; therefore, the magnitude of 
$R_{\rm long}$ (longitudinal HBT radius) 
is proportional to the total duration of the longitudinal expansion, 
i.e. to the decoupling time of the system.
The decoupling times extracted from this fit to the ALICE radii and to the values published at lower energies are shown in Fig.~\ref{fig:hbt} (right). As can be seen, $\tau_f$ scales with the cube root of the charged-particle multiplicity and reaches 10--11~fm/$c$ in central Pb--Pb collisions at the LHC.

\section{Suppression of high-$\pt$ charged particle production: an intriguing pattern}
\label{sec:raa}

\begin{figure}[t!]
\begin{center}
\includegraphics[width=0.5\textwidth]{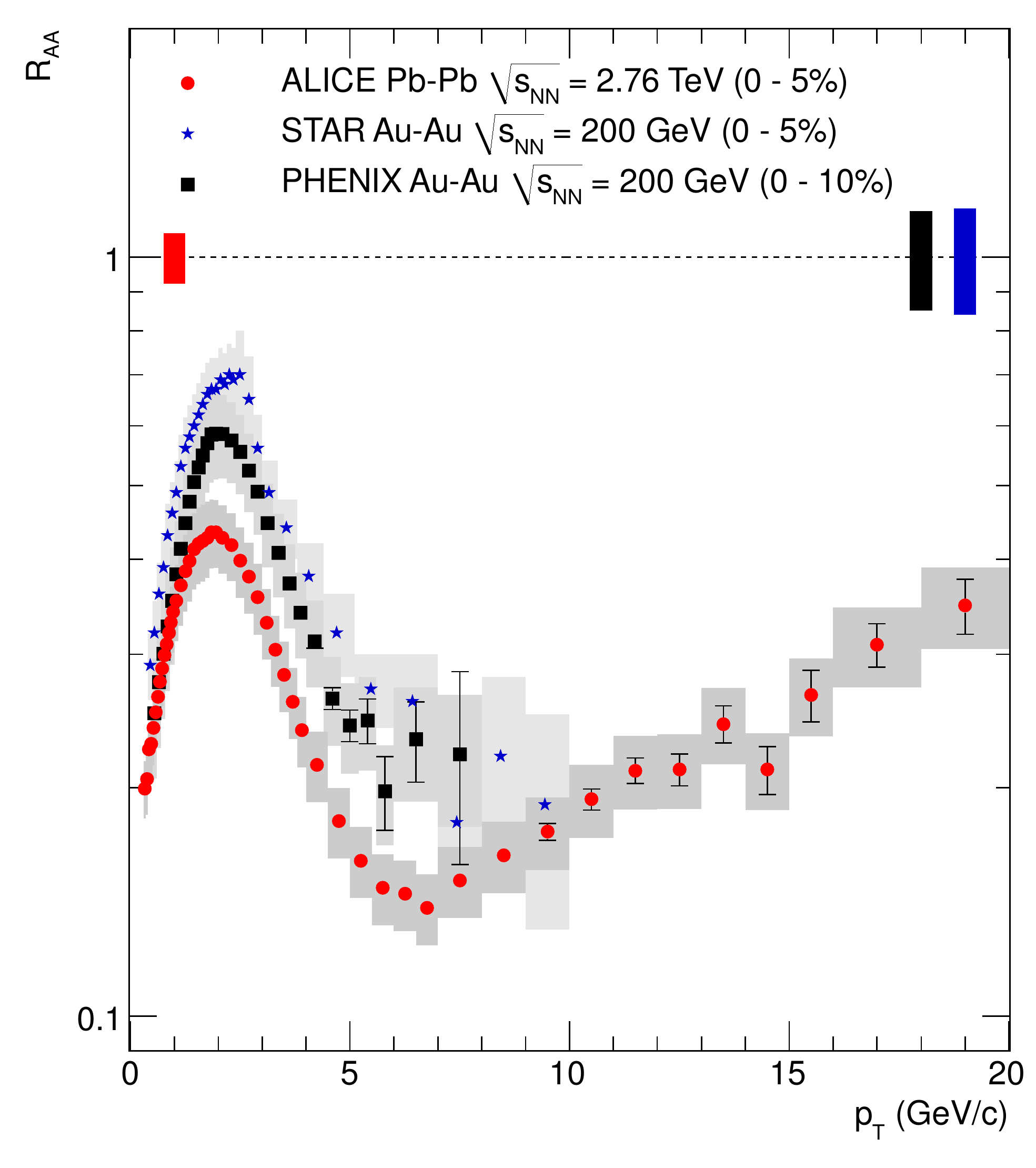}
\caption{
Nuclear modification factor $R_{\rm AA}$ for charged hadrons 
in central Pb--Pb collisions at the LHC, compared
 to measurements at 
$\sqrtsNN=200$~GeV by the PHENIX and STAR experiments. 
%The statistical and systematic errors of the ALICE and PHENIX data are shown as error bars and boxes, respectively. The statistical and systematic errors of the STAR data are combined and shown as boxes. The vertical bars around $R_{\rm AA} = 1$ indicate the $\pt$ independent scaling errors.
See~\cite{raa} and references therein.}
\label{fig:raa} 
\end{center}
\end{figure}

One of the most awaited-for  measurements in heavy-ion collisions at the LHC 
is certainly the nuclear modification factor of charged hadrons, which ten years ago
at RHIC yielded the first indication of the jet quenching phenomenon, now commonly 
attributed to parton energy loss in hot and dense QCD matter. This observable is defined
as $R_{\rm AA}(\pt) = ({\rm d}N_{\rm AA}/{\rm d}\pt)/(1/\langle N_{\rm coll}\rangle \, {\rm d}N_{\rm pp}/{\rm d}\pt )$, that is, 
the ratio of the $\pt$ spectrum measured in nucleus--nucleus to that expected on the basis of the proton--proton spectrum scaled by the number $N_{\rm coll}$ of binary 
nucleon--nucleon collisions in the nucleus--nucleus collision (as calculated in the Glauber model).
At RHIC energies, 
the $R_{\rm AA}$ factor was measured to be of about 0.2 and roughly independent of $\pt$
in the range 5--15~GeV/$c$, i.e. a factor of five suppression 
in high-$\pt$ particle production with respect to pp collisions.

The charged particles $R_{\rm AA}$ was measured by ALICE out to $\pt=20$~GeV/$c$
after a few days of the end of the Pb--Pb run~\cite{raa}.
Charged particle tracks were reconstructed using information from the TPC and ITS detector systems in the region $|\eta|<0.8$. 
The primary track selection described in section~\ref{sec:flow} was applied.
When $R_{\rm AA}$ was first evaluated, no measured pp reference at
 $\sqrts = 2.76$~TeV existed\footnote{The LHC was run with pp collisions at this energy 
 later, in March 2011.}. As explained in~\cite{raa}, a reference was constructed by 
 interpolating the ALICE measurements at $\sqrts=0.9$ and 7~TeV.
Figure~\ref{fig:raa} shows the nuclear modification factor $R_{\rm AA}$ of charged hadrons
for central Pb--Pb collisions, compared to the same measurement by the PHENIX and STAR
experiments at RHIC.  
In central collisions at the LHC, $R_{\rm AA}$ exhibits a very strong suppression, 
reaching a minimum of $\approx 0.14$ at $\pt= 6$--7~GeV/$c$. 
Despite the much flatter $\pt$ spectrum in pp at the LHC, the nuclear modification factor 
at $\pt = 6$--7~GeV/$c$ is smaller than at RHIC. This suggests an enhanced energy loss at LHC and therefore a denser medium. 
A significant rise by about a factor of two is observed for $7 < \pt < 20$~GeV/$c$. This pattern 
is very intriguing, because it suggests that very high momentum partons may lose only a small
fraction of their energy in the medium and, thus, be sensitive probes of its properties.
%A quantitative determination of the energy loss and medium density will require further investigation of gluon shadowing and saturation in the present energy range and detailed theoretical modeling.

\section{Ongoing analyses}
\label{sec:outlook}

\begin{figure}[t!]
\begin{center}
\includegraphics[width=0.46\textwidth]{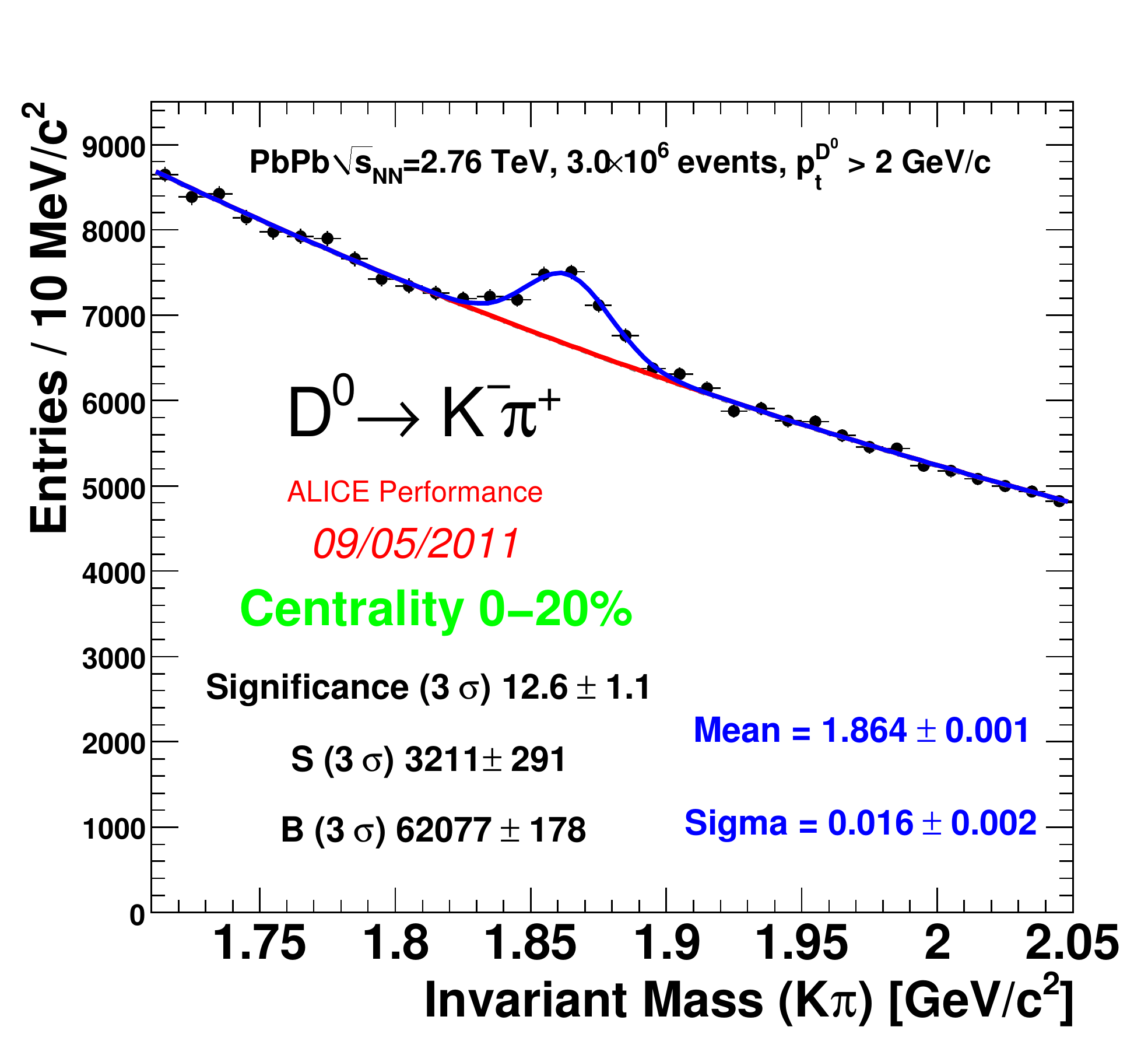}
\includegraphics[width=0.53\textwidth]{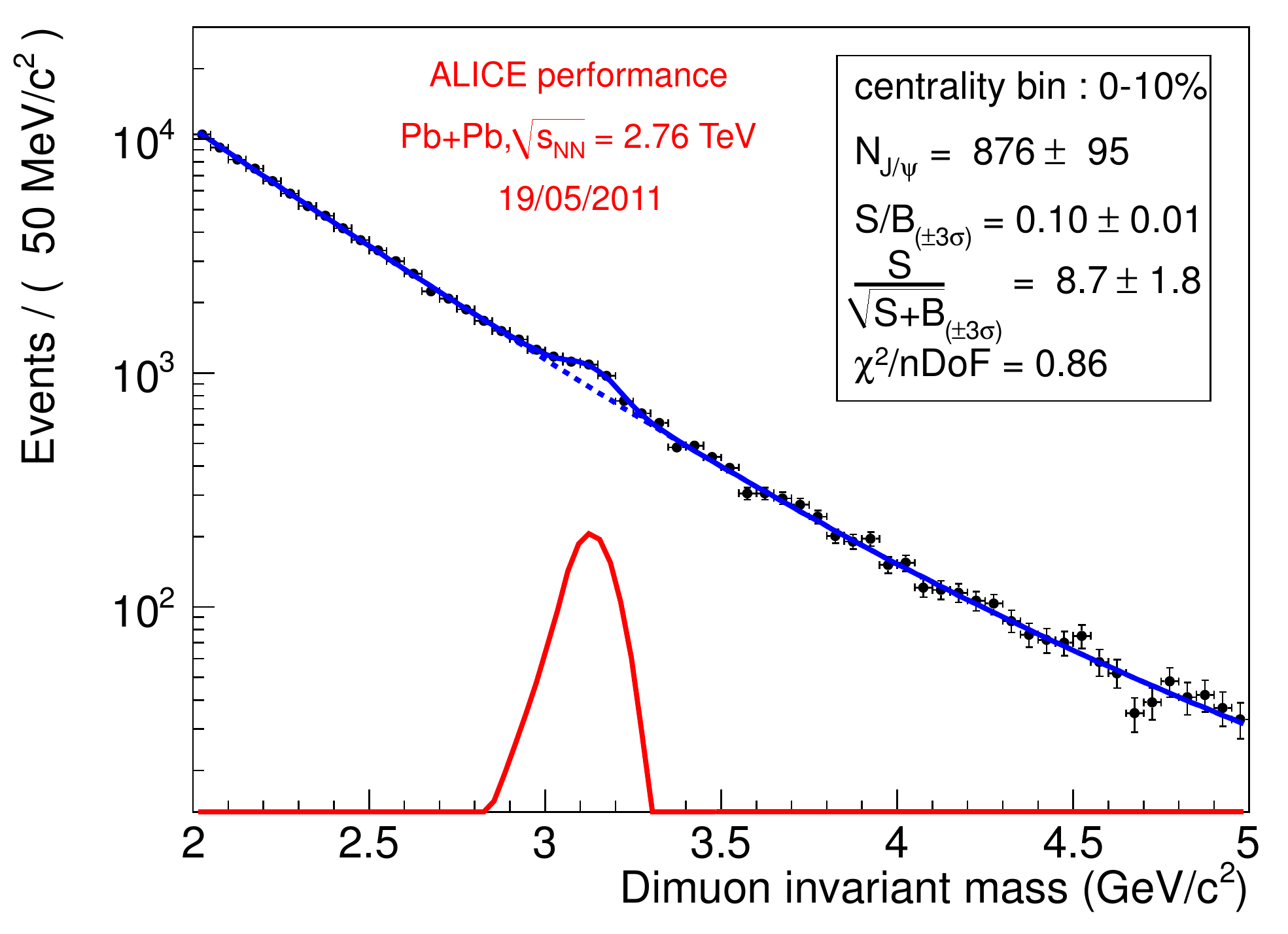}
\caption{Signals of charm particles in central Pb--Pb collisions at the LHC.
Left: $\rm D^0\to K^-\pi^+$ at central rapidity ($|y|<0.8$). Right: 
$\rm J/\psi\to \mu^+\mu^-$  at forward rapidity ($2.5<y<4$).}
\label{fig:charm} 
\end{center}
\end{figure}

Several measurements of strange and heavy-flavour particle production 
have been carried out, as well as studies of jet production and jet-like particle 
correlations~\cite{qm}. As examples of the ALICE detector performance in Pb--Pb
collisions, we report in Fig.~\ref{fig:charm} the invariant mass distributions showing the
signal of $\rm D^0\to K^-\pi^+$ 
decays selected using displaced decay vertices at central rapidity and 
$\rm J/\psi\to \mu^+\mu^-$ decays in the forward muon spectrometer.

\section{Conclusions}
\label{sec:concl}

We have presented the first ALICE physics results from Pb--Pb collisions at the LHC.
\begin{itemize}
\item The highest ever reached charged-particle multiplicity was measured, suggesting
that a system with an energy density at least three times higher than at RHIC energies
is produced. 
\item The volume of the particle emitting source is found to be twice larger than at RHIC 
energies.
\item The collision-centrality dependence of the particle multiplicity tends to flatten towards
most central collisions, suggesting that some kind of saturation mechanism is at play for 
the initial-state gluon fields in the colliding nuclei.
\item The produced hadrons exhibit a strong collective flow, in agreement with the hydrodynamic
 models  
in which the system expands similarly to a liquid with very small viscosity.
\item High-momentum particle production shows a suppression by a factor of 5--7, 
close to that observed at lower energy in the same momentum range, suggesting 
that the medium opaqueness to hard partons is higher at the LHC.
\end{itemize}
The study of these observables for many species of identified particles 
(baryons, strange, heavy-flavour particles, charmonia) is well-advanced and 
opens the path for a detailed characterization of the Quark-Gluon Plasma state produced 
in the highest-energy nuclear collisions at the LHC.

\end{document}